\documentclass[11pt]{article}

\usepackage[margin=1in]{geometry}

\usepackage{enumerate}

\usepackage{tikz}
\usetikzlibrary{shapes,arrows}
\usepackage{amsthm,amsfonts,mathtools}
\usepackage{amsmath}
\usepackage{amssymb}
\usepackage{latexsym}
\usepackage{verbatim}
\usepackage{xcolor}
\usepackage{booktabs}

\usepackage{multirow}
\usepackage{mathrsfs}
\usepackage{graphicx}
\usepackage{subfigure}
\usepackage{hyperref}
\usepackage{bbm}
\usepackage{dsfont}
\usepackage{tipa,bm}
\usepackage[linesnumbered,vlined,ruled]{algorithm2e}
\usepackage{textcomp}
\usepackage{soul}

\usepackage[numbers]{natbib}

\DeclareMathOperator{\var}{\text{var}}

\newcommand{\E}{\mathbb{E}}

\newcommand{\cF}{\mathcal{F}}

\renewcommand{\P}{\mathbb{P}}

\definecolor{blue1}{RGB}{205,240,255}
\definecolor{blue2}{RGB}{150,210,255}
\definecolor{yellow1}{RGB}{255,253,210}
\definecolor{yellow2}{RGB}{255,250,190}
\definecolor{red1}{RGB}{255,220,220}
\definecolor{red2}{RGB}{255,200,200}
\definecolor{red3}{RGB}{255,50,50}
\definecolor{red4}{RGB}{0,0,0}
\definecolor{white}{RGB}{255, 255, 255}

\title{European Football Player Valuation: Integrating Financial Models and Network Theory}

\author{
  \and  Albert Cohen \\
  Department of Mathematics \\
  Department of Statistics and Probability \\
  Michigan State University \\
  East Lansing, MI 48824 \\
  \texttt{acohen@msu.edu}
  \and  Jimmy Risk \\
  Mathematics \& Statistics \\
  Cal Poly Pomona \\
  Pomona CA 91676\\
  \texttt{jrisk@cpp.edu} \\  
}

\begin{document}

\maketitle

\begin{abstract}
This paper presents a new framework for player valuation in European football, by fusing principles from financial mathematics and network theory. The valuation model leverages a "passing matrix" to encapsulate player interactions on the field, utilizing centrality measures to quantify individual influence. Unlike traditional approaches, such as regressing on past performance-salary data, this model focuses on in-game performance as a player's contributions evolve over time.  Consequently, our model provides a dynamic and individualized framework for ascertaining a player's fair market value. The methodology is empirically validated through a case study in European football, employing real-world match and financial data. This cross-disciplinary mechanism for player valuation adapts the effect of connecting pay with performance, first seen in Scully \cite{Scully}, to include in-game contributions as well as expected present valuation of stochastic variables.
\end{abstract}


\section{Introduction}

Player valuation in European football is a complex endeavor that necessitates nuanced metrics beyond traditional sports statistics.  Classical economic theory suggests that a profit-maximizing firm will hire by setting wage equal to marginal revenue product (MRP) of labor, that is, the contribution to revenue by a unit of labor \cite{RockerbieEaston}.  The first major work to address the link between performance and pay for athletes is by Scully \cite{Scully}, widely recognized as one of the founders of player compensation and free agency theory in sports. Scully proposed a marginal revenue product (MRP) theory for player pay utilizing well-known tools in labor economics, linking on-field play with salary for baseball players. A player should be compensated in relation to their contribution to team performance. Specifically, the player should be paid relative to each extra dollar of revenue they generate due to their performance on the team, related to their marginal effect on winning percentage \cite{rockerbie2010marginal}.

This approach is valuable in setting a structural foundation for linking contract value with performance. There is in fact a deep literature regarding the work of Scully and subsequent authors, and the reader is referred to the works of \cite{brown2009impact,leeds2001winner,krautmann1999s,macdonald1994baseball,zimbalist2001salaries,zimbalist2010reflections}.  

Although foundational and readily applicable, evidence suggests that a simple ``MRP = Wage'' is not fully sufficient for player valuation.  For example, Rockerbie and Easton \cite{RockerbieEaston}[Chapter 2] finds that in baseball, higher-priced free agents are overpaid relative to their MRP, while lower-paid free agents offer a bargain.  The cause for this could be for a variety of reasons, one being that paying MRP as wage holds true when teams are profit-maximizing \cite{rosen2001labour}, but not necessarily when teams are win-maximizing. Kesenne \cite{kesenne2000revenue} find that in European football clubs, win-maximizers can result in different revenues and payroll than under profit maximization.  Another is that sports employment contracts are typically multi-year with a fixed payment schedule, failing to take into account a multitude of stochastic fluctuations including economic conditions, player injuries, and general player performance. 




\subsection{Stochastic Approaches to Player Valuation}

The work of Rockerbie and Easton \cite{RockerbieEaston} in their recent book offers a discrete, multi-period approach to contract pricing via expected present valuation, addressing the limitations of the classical economic model that assumes labor is hired until the marginal revenue product of labor just equals the wage rate. They state that this classical approach, being static, fails to account for market fluctuations and varied worker productivity. In contrast, Rockerbie and Easton provide a dynamic framework that allows teams to adapt their strategies based on emerging market conditions and individual player performance, offering a practical tool for valuing players. Their approach determines salary values by modeling player performance as one of three possible outcomes each season—exceeding, meeting, or falling short of preseason expectations. This trinomial model directly links the dollar amount of the marginal revenue product (MRP) to performance in each of these outcomes, enabling more accurate decision-making under uncertainty, such as the option to re-sign a player.

From an alternative perspective, Tunaru, Clark, and Viney (TCV) \cite{Tunaru}, address the value of a player to a European Football team, via continuous-time stochastic modeling and a subsequent Black-Scholes partial differential equation (PDE). There have been some numerical implementations of this model, such as in Coluccia et al.\cite{coluccia2018application}~to value the goalkeeper of a Serie A League club. 

At its core, the TCV model uses a given \emph{player performance index} to which contingent claims style modeling is used to develop the value of the player. In particular, they use the Opta Index \cite{optadefn} as an index of player performance, which is assumed to follow a Geometric Brownian Motion (GBM) \cite{shreve2004stochastic}, a common assumption in contingent claims modeling.  Team revenue and the sum of player indices to determine revenue per index point are also modeled by GBM's, in which case a single player's value can be determined by multiplying this ratio with their index.


\subsection{Indices of Player Performance}

It is obviously difficult to provide a single number to fully summarize a player's in-game contributions. As mentioned above, the Opta Index \cite{optadefn} is one way of attempting this in European football. The Opta Points system evaluates a player's overall contribution based on a linear combination of individual performance metrics such as goals, assists, passes, tackles, interceptions, and more. According to TCV \cite{Tunaru} (in 2006), the index is calculated from the Points of the last 6 games, with the minutes of gameplay kept as an exposure factor.  The exact coefficients used to compute a player's Opta point value are available on the Opta website \cite{opta} at the time of this writing; however, up-to-date information on how to compute the index from the Points, or historic indices of players, could not be found.

It is worth noting that these Points are formed as a linear combination of marginal post-game player stats.  An alternative is to build directly on the most fundamental aspect of European football, which is the network formed from the ball transitioning from player to player or goal.  In a given game or season, one can construct a \emph{passing matrix} whose $i,j$ entry is the number of passes from player $i$ to player $j$.  By row-normalizing this matrix so that each row sums to 1, it provides the \emph{proportion of passes} from player $i$ to player $j$, or equivalently the \emph{empirical probability} of that transition.  Matrix centrality measures applied to passing matrices, as studied by Pena et al.~\cite{pena2012network}, capture a player's importance in play development and can be used to provide a single value to quantify their performance for that game or season.  As not all passes are equal--some are more likely to lead to scoring or turnover chances--Duch et al.~\cite{duch2010quantifying} extends this to an \emph{augmented passing matrix} that includes shots on goal. By row-normalizing, these matrices become Markov chains, allowing relevant theory to be used.  As an example, using this matrix, it is straightforward to compute the probability that any given player is involved in a team possession that ended in a goal.  This type of metric was used in Duch et al.~\cite{duch2010quantifying} and provides value to players that are crucial to a scoring play but may be underappreciated by post-game summary statistics.  Additionally, defensive contributions naturally embed themselves as the initial distribution of the Markov chain, that is, the probabilities of players beginning possessions.

\subsection{Our Contributions}

The novelty in our work is two-fold. First, we develop a financial model for player performance that improves upon the TCV approach \cite{Tunaru}. We directly model the proportion of a player's index relative to the sum of all players' indices.  This proportion actually appears directly in the underlying TCV model, but their work models the numerator and denominator separately with a simplifying assumption that the sum of player indices (each a GBM), despite not being a GBM, effectively constitutes a GBM for modeling purposes. Directly modeling the proportion of player performance to the whole team avoids this assumption. Additionally, the model includes a parameter for revenue share that goes to team salary.  This is an improvement over TCV whose consequent player value is not directly comparable with salary.  Including this as a known parameter allows our model to hold regardless of whether a club is win- or profit-maximizing, as this is encoded in the proportion of salary that goes out to players.  Thereafter, we combine elements from TCV and Rockerbie and Easton \cite{RockerbieEaston} to determine multi-year contract values. This combined method aligns with the contingent claims literature and provides a dynamic tool for attaching financial value to player performance. Consequently, clubs can better assess contract extensions, midseason trades, or the release of players whose performance does not justify their salaries. 

Second, we rigorously discuss the passing matrix and its augmented version as a Markov chain.  This approach provides closed formulae for quantities that may be useful to practitioners, and contains existing methodologies, like the flow centrality measure by \cite{duch2010quantifying}.  We find that an extension of this measure provides a suitable index that can be modeled and used for player valuation.  

The rest of the paper is organized as follows.  In Section \ref{sec:stoch-models}, we develop the stochastic model for player valuation that incorporates both the uncertainty in player performance and the economic conditions affecting a team's revenue. Exact contract pricing is discussed, which balances the expected present value of a player's stochastic performance-linked contributions against their fixed salary, allowing for dynamic contract evaluation throughout the term. The specific process used for player performance is detailed in Section \ref{sec:passing-matrix}, where (augmented) passing matrices are discussed as Markov chains, with an example provided to illustrate how player role affects their performance value.  Next, Section \ref{sec:full-model-usage} provides a template for how the entire valuation model can be used from start to end, and applies it to an example to illustrate its flexibility.  Section \ref{sec:case-study} showcases a full case study on nine players in the English Premier League (EPL) over five seasons (2018--2023).  This case study analyzes the fitted performance index model for each player and subsequently compares their valuation to actual salary over the period.  Finally, Section \ref{sec:conclusion} concludes and discusses future work and limitations.

\section{Stochastic Models for Player Valuation}\label{sec:stoch-models}

Consider a scenario where a player can sign a contract with a team for multiple years. There is a baseline expectation for performance, but management allows for the possibility that the player can under- or over-perform, in terms of expected statistical contribution. 
To set ideas, subscripts $i$ and $j$ always refer to a player, and $t$ to a time.  Our model for the stochastic value of player $j$ at time $t$ (in years) is 

\begin{equation} \label{Yextension}
    Y_{j,t} = \pi_{j,t} a_t R_{t},
\end{equation}

\noindent where 

\begin{itemize}
    \item $a_t$ is the \emph{proportion of team revenue that goes to players} at time $t$.
    \item $\pi_{j,t} \in (0,1)$ is player $j$'s \emph{team-standardized performance index} at time $t$, and 
    \item $R_{t}$ is the revenue at time $t$.
\end{itemize}

\noindent Although it technically is not needed in the math that follows, a useful assumption is that $\sum_j \pi_{j,t} = 1$, allowing $\pi_{j,t}$ to be thought of as a proportion of player performance relative to the entire team.  This also recovers value paid to the entire team as $\sum_j Y_{j,t} = a_t R_t$.  

\emph{Remark.} It is worth noting that TCV \cite{Tunaru} models the value of a player as
\[ \frac{R_t}{\sum_i q_{i,t}} \cdot q_{j,t} = R_t \pi_{j,t}, \qquad \pi_{j,t} = \frac{q_{j,t}}{\sum_i q_{i,t}},\]

\noindent where $q_{j,t}$ is the \emph{non-standardized index point value} of player $j$ at time $t$.  The reason for writing it this way is because $R_t / \sum_i q_{i,t}$ is value in money for a single point of the team.  This value is not standardized by $a_t$ and so is not usable as a surrogate for a salary.  In what follows we assume $a_t$ is known ahead of time, effectively encoding information about whether the club is a win-maximizer (larger $a_t$) or profit-maximizer (smaller $a_t$).

Now, a deterministic salary stream must be determined at the time of contract negotiations prior to playing.  The foundational idea in contingent claim valuation is the notion of swapping a fixed payment or series of payments for a random payment or random series (see \cite{hull2016options} for a more comprehensive reference). This is of course a non-trivial task, as an athlete would like a predetermined salary, yet there is risk of both injury and under-performing expected on-field contributions, and so the question arises of how to properly average the possible dollar amounts corresponding to how the season may play out for the player and their team. This is the idea driving our approach to valuing in-game performance as a random dollar value exchanged for a certain salary value agreed to by a player or their agent. If the two sides do not match, or are at least not relatively close, either party to negotiations may determine the player's contract is due for review.

\subsection{Player Value Swaps}

Assume player $j$ is fixed.  For brevity, we drop the subscript in what follows except in cases where its importance is elaborated.  Denote

\begin{itemize}
    \item $r$ is the \emph{risk-free rate}, a guaranteed rate of return compounded continuously on an alternative market investment,
    \item $\lambda$ is the \emph{risk-premium} for the player,
    \item and $C_{t}$ is the performance-linked salary fixed at the time of contract negotiations to be paid to the player at time $t$.
\end{itemize}  

Here, $t$ is in years but depending on context could be considered as an annual salary $t=1, 2, \ldots$ or weekly salary $t=1/52, 2/52, \ldots$. Elaborating through a one-year contract example, a team owner can always invest a rate of $r$ to pay $C_{1}$ at the end of year 1, meaning the \emph{present value} of this deterministic payment is $C_{1} e^{-r}$.  On the other hand, the stochastic value of the player at the end of year 1 is expected to be $\E[Y_{1}]$.  The \emph{risk-averse investor} recognizes the risk involved, since an investment of $e^{-r}\E[Y_{1}]$ is not guaranteed to accrue to the stochastic $Y_{1}$ like in the risk-free case. The risk-premium $\lambda$ is an additional discount rate that makes the investment worthwhile.  In particular, it makes the expected present values equal, that is,

\begin{equation}\label{eq:yr1-salary-balance}
    C_{1} e^{-r} = e^{-(r+\lambda)}\E[Y_{1}].
\end{equation}

More generally, we have for a $T-$year contract:

\begin{itemize}
    \item the negotiated salary payments, $C_{1},C_{2},...,C_{T}$ with the amount $C_{t}$ computed as if paid at the end of the $t^{th}$ year, and
    \item the random dollar-value of the athlete's performance-linked contributions on the field, $Y_{1},Y_{2},...,Y_{T}$.
\end{itemize}  

We seek to find a balance, or \emph{swap}, of the fixed salaries for the unknown performance values. Applying the same principles that were used to derive \eqref{eq:yr1-salary-balance}, the balance is now achieved by setting the expected present value of payment schemes to be equal, that is, 

\begin{equation} \label{balance}
\sum_{t=1}^{T} e^{-r t}C_{t} =  \sum_{t=1}^{T} e^{-(r+\lambda) t} \E[Y_{t}].
\end{equation}









With Equation \eqref{balance}, salaries can be constant throughout the term, be front-loaded, or have other term-structures that match the expected present value linked to revenue streams on the right side of \eqref{balance}. This balance equation provides a way for general managers to determine a player's contract structure before the season begins. For that reason, we assume $a_t$ is known ahead of time to both parties.

\subsection{Midterm Valuation}

A question that naturally arises in a manager's mind is how a player's value during the term of the contract compares to their fixed salary value. Thus, it is useful to consider a dynamic measure of this gap between these two valuations of a player's contributions on the field. Comparing $C_{t}$ directly to the stochastic $Y_{t}$ is not ideal since it does not take into consideration future uncertainties.
In order to see how a contract negotiated at time $0$ has aged by the middle of the term, we define the time $t$ valuation of a player $j's$ performance-linked value as $S_{t,t+h}$. In words, this is the new (rolling) value of a fixed contract if the manager could re-negotiate with a single payment at time $t$ and single stochastic valuation at time $t+h$. Using the same principles as in Equation \eqref{eq:yr1-salary-balance}, the agreed upon time $t$ value at time $t+h$ is $S_{t,t+h}$, where

\begin{equation}\label{eq:conditional-valuation}
    e^{-rh} S_{t,t+h} = e^{-(r+\lambda)h} \E[Y_{t+h} | \mathcal{F}_{t}].
\end{equation}

\noindent Here, $\mathcal{F}_{t}$ refers to the information available up to time $t$, i.e.~$R_t \in \mathcal{F}_t$ and $\pi_t \in \mathcal{F}_t$ for all $t \geq 0$, so that a \emph{conditional expectation} can be taken with respect to that information.   This is made rigorous through the concept of a \emph{filtration} $\mathcal{F} = (\mathcal{F}_t)_{t \geq 0}$, the collection of all information sets. We provide additional details in Appendix \ref{app:filtration}, including some formulae we will use in the sequel (also, see \cite{shreve2004stochastic}).

Equation \eqref{eq:conditional-valuation} can be further generalized to be consistent with a multi-year forward-looking valuation at time $t$ as in \eqref{balance} noting that valuation happens at time $t$ and should only include information up to that point, as well as only including terms for future times.  In particular,

\begin{equation}\label{eq:balance-look-ahead}
    \sum_{k > t}^{T} e^{-r(k-t)} C_{t,k} = \sum_{k>t}^T e^{-r(k-t)} S_{t,k}.
\end{equation}

\noindent where $C_{t,k}$ is the salary for the player at time $k$ decided at time $t$, and $k$ spans the valuation times after $t$.  Note that the initial salary structure as in Equation \eqref{balance} can be recovered by setting $t=0$.



\subsection{Dynamics of Player Value Processes}

We begin the process by modeling on-field performance as stochastic, due to innate ability, matchups, injury, etc. Consider the \emph{univariate approach} of a single player.  The model for $Y$ is presented in continuous time for flexibility. We assume the reader is familiar with these types of continuous-time models; see \cite{shreve2004stochastic} or \cite{revuz2013continuous} for details.  
In particular, the full dynamics are defined as

\begin{equation} \label{sde1}
    \begin{aligned}
        Y_{t} & = \pi_{t} a_t R_{t}  \\
       d \pi_{t} & = -\theta(\pi_t - \pi^*)dt + \sigma_{\pi} \sqrt{ \pi_{t} (1-\pi_{t})} dW^{\pi}_{t} \\
       dR_{t} & = \mu R_{t} dt + \sigma_{R} R_{t} dW^{R}_{t} \\
       \rho dt & = \langle dW^{\pi}_{t},dW^{R}_{t}\rangle, \\
    \end{aligned}
\end{equation}

\noindent where $\sigma_R, \sigma_\pi, \theta > 0$, $-1 \leq \rho \leq 1$, $0 \leq \pi^* \leq 1$, and $\min(\pi^*, 1-\pi^*) \geq \sigma^2_\pi / (2\theta)$.  Here, $W^\pi$ and $W^R$ are Brownian motions \cite{shreve2004stochastic} with correlation $\rho$ as indicated in the last line of \eqref{sde1}.  Note that $R$ follows a geometric Brownian motion (GBM) \cite{shreve2004stochastic} with growth rate $\mu$ and volatility $\sigma_R$.  The process defined by $\pi_t$ lies in the family of Pearson diffusion processes  \cite{forman2008pearson}.  These have well known properties.  To name a few, $\pi_t$ is guaranteed to be bounded on $(0,1)$ and $\pi_t$ has stationary (long-term) distribution Beta with known parameters in terms of $\theta, \pi^*$, and $\sigma_\pi$ \cite{bakosi2010exploring}.  The parameter $\theta$ controls the rate of reversion to the stationary mean $\pi^*$, which can be seen through \eqref{sde1}.  Consequently, 
\begin{equation}\label{eq:pi-moments}
\E[\pi_t] = \pi^*, \qquad \var(\pi_t) = \pi^*(1-\pi^*)\frac{\sigma_\pi^2}{2\theta + \sigma_\pi^2}, \qquad \varrho_\pi(h) = \exp(-\theta h),  
\end{equation}
where $\E[\pi_t]$ and $\var(\pi_t)$ are the \emph{stationary (long-run) mean} and \emph{variance} of $\pi_t$ and $\varrho_\pi(h) = \text{corr}(\pi_{t}, \pi_{t+h})$ is the \emph{autocorrelation} of $\pi$ across $h$ time units \cite{bakosi2010exploring, forman2008pearson}.  It is possible for $\theta, \pi^*$, and $\sigma_\pi$ to be time-varying; see \cite{bakosi2010exploring} for details.





Toward finding $S_{t,t+h}$, the performance value at time $t$ for period $t+h$, assume the payment structure is known in the future so that $a_{t+h} \in \cF_t$.  Using Equation \eqref{eq:conditional-valuation},

\begin{equation*}
    S_{t,t+h} = e^{-\lambda h} \E[Y_{t+h} | \cF_t] = a_{t+h} e^{-\lambda h} \E[\pi_{t+h} R_{t+h} | \cF_t]. 
\end{equation*}

Computing $\E[\pi_{t+h} R_{t+h} | \cF_t]$ requires using the dynamics in \eqref{sde1} and is difficult due to the correlation between $dW_t^R$ and $dW_t^\pi$.  We use a first order Taylor series approximation of  $v(\pi_t) = \sqrt{\pi_{t} (1-\pi_{t})}$ about its mean reversion parameter $\pi^*:$ 

\begin{equation}\label{eq:taylor-approx}
v(\pi_t) = v(\pi^*) + v'(\pi^*)(\pi_t - \pi^*) + \mathcal{O}((\pi_t - \pi^*)^2), \qquad v'(\pi^*) = \frac{1-2\pi^*}{2\sqrt{\pi^*(1-\pi^*)}}.
\end{equation}

\noindent This approximation is justified since $\pi_t$ is mean-reverting to $\pi^*$.  Truncating the $\mathcal{O}((\pi_t-\pi^*)^2)$ terms, we obtain the first order approximation





\begin{equation} \label{ctssalary-solved}
S_{t,t+h} =e^{(\mu-\lambda)h} a_{t+h} R_t \left[\pi^* + (\pi_t - \pi^*) e^{(cv'-\theta) h} + c \sqrt{\pi^*(1-\pi^*)}\frac{1-e^{(cv'-\theta) h}}{\theta-cv'}\right],
\end{equation}

\noindent where $c=\rho\sigma_R\sigma_\pi$ is a covariance term and we use the shorthand $v' := v'(\pi^*)$ as in \eqref{eq:taylor-approx}.  Appendix \ref{app:salary-calculation} details the derivation.  

The result in \eqref{ctssalary-solved} has some intuitive properties consisting of two multiplicative terms.  The first is $e^{(\mu-\lambda)h} a_{t+h} R_t$, which is the current revenue $R_t$, along with its projected growth by $\mu$ by $h$ years, adjusted by the proportion of revenue $a_{t+h}$ paid to players at time $t+h$.  This can be thought of as the team's pay at time $t+h$, which is finally discounted by the player's risk premium $\lambda$.  The latter is a term related to the player's share relative to the team, and is the sum of three individual terms:

\begin{itemize}
    \item $\pi^*$: The long-run expectation of the player's proportional contribution to the team.  If $\pi_t$ was not stochastic and $\pi^*$ was used instead, the other two terms become zero. 
    \item $(\pi_t - \pi^*) e^{(cv'-\theta) h}$: This is related to mean reversion, as the player's current performance is $\pi_t$.  For example, if the player is currently underperforming with respect to their anticipated long-term $\pi^*$, then $\pi_t - \pi^* < 0$, negatively contributing to their value.  This is reasonable when a player is in a slump.  Note that this decays according to the rate $\theta - cv'$.  Indeed, over longer horizons ($h$), their current performance $\pi_t$ becomes less impactful as they will tend to their long-term average $\pi^*$.  The rate of decay naturally contains the mean-reversion parameter $\theta$, adjusted by $cv'$.  Since $v' = \frac{1-2\pi^*}{2\sqrt{\pi^*(1-\pi^*)}} > 0$ (assuming $\pi^* < 0.5$), the volatility is an increasing function, meaning this adjustment slows down the mean reversion by $cv' = \rho \sigma_R \sigma_\pi \frac{1-2\pi^*}{2\sqrt{\pi^*(1-\pi^*)}}$ when $\rho > 0$; that is, the stochastic shocks impacting $R_t$ ``interfere" with the mean-reversion of $\pi_t$.
    \item $c \sqrt{\pi^*(1-\pi^*)}\frac{1-e^{(cv'-\theta) h}}{\theta-cv'}$: This is a covariance term between revenue and performance that provides value according to the player's stochastic performance.  From a contingent claims perspective, this is the present value of $c \sqrt{\pi^*(1-\pi^*)}$ accruing continuously at a rate of $\theta - cv'$ for $h$ years \cite{wilders2020financial} (this rate was discussed in the last bullet point).  Smaller mean-reversion rates ($\theta$) increase this value.  In essence, covariation contributes positively to pay (when $\rho > 0$), but not indefinitely due to the mean reversion.
\end{itemize}

We discuss two ways the expression for $S_{t,t+h}$ in Equation \eqref{ctssalary-solved} can be utilized.  Firstly, it can be used directly for full contract valuation with the balance Equation \eqref{eq:balance-look-ahead}; this is explored in Section \ref{sec:illustrative-example-valuation}.  Second, one can consider the process $(S_{t,T})$ for $t \leq T$ which represents the player value process's evolution in $t$ for a fixed terminal time $T$.  In this case, a useful measure is
\begin{equation}
    \Delta_{t,T}(C) = C - S_{t,T}, \qquad t \leq T \label{eq:delta}\\
\end{equation}
where $C$ is a baseline value of performance, commonly set to a fixed salary when there is a uniform structure.  This depends on $\pi_t$ and $R_t$ through $S_{t,T}$ and thus has its own dynamics according to \eqref{sde1}.  This offers a running measure of comparative value, since $\Delta_{t,T}(C) > 0$ means $S_{t,T} < C$, suggesting underperformance relative to $C$.  Similarly $\Delta_{t,T} < 0$ suggests overperformance (i.e.~$S_{t,T} > C)$. Note that it has its own variability and should not inform immediate decision making like a player transfer, but  could be further studied as the basis for insurance against overpayment (or underpayment) in a fixed contract.  For example, if it widens beyond a certain threshold, the manager may be tempted to release or trade the player during the upcoming season.  It could also be used to include add-ons into a player's contract, say, a bonus if the average $\Delta_{t,1}(C)$ over season 1 was negative.  Such a bonus could be priced at time $t=0$ using the arguments in Section \ref{sec:stoch-models} according to a payoff of $\max(0, \int_0^1 S_{t,T}dt-C)$, with dynamics described by \eqref{ctssalary-solved} and \eqref{sde1}.

We end this section with a few remarks.  

\begin{itemize}
    \item Appendix \ref{app:salary-alternative} highlights alternative ways to solve for $\E[\pi_{t+h} R_{t+h} | \cF_t]$ using the dynamics in Equation \eqref{sde1} including Monte Carlo simulation \cite{glasserman2004monte} to obtain $\E[\pi_{t+h} R_{t+h} | \cF_t]$ as well as a partial differential equation approach using the Feynman-Kac \cite{shreve2004stochastic} formula that can be solved numerically.  The appendix highlights a numerical example that illustrates effectiveness of the first order approximation.
    \item In the case of $\rho=0$, $S_{t,t+h}$ in Equation \eqref{ctssalary-solved} is exact.  In particular, $R_{t}$ and $\pi_{t}$ are now uncorrelated and both $\E[Y_{t+h} | \cF_t]$ and $\E[\pi_{t+h} | \cF_t]$ are known.
    \item The above discussion assumes $\pi^* < 0.5$ and $\theta > cv'$.  The former assumption is always satisfied in practicality, since a player's long-run performance will never hold over 50\% of that of the team (otherwise, the mean-reversion is accelerated).  The latter is also satisfied in practicality since $\min(\pi^*, 1-\pi^*) \geq \sigma^2\pi / (2\theta)$.
\end{itemize}

\section{Model for Player Performance}\label{sec:passing-matrix}

In theory, $\pi_t$ can be represented by any player performance metric bounded between 0 and 1.  Our method caters to the specifics of team sports that crucially rely on frequent changes of possession and utilizes augmented passing matrices. It is closely related to the flow centrality measure \cite{duch2010quantifying} and is made rigorous here.




Consider the \emph{passing matrix} for a team of $M$ players, which is an $M \times M$ matrix whose $i,j$ entry is the probability that player $i$ passes to player $j$ when player $i$ has the ball.  This could be estimated from empirical data (standardizing so that all rows sum to one), and/or utilize expert opinion. To value scoring attempts and devalue missed passes and turnovers, we augment the passing matrix with two rows and columns: one for shots ($S$) and another for unsuccessful passes and/or turnovers ($U$).  Denote this \emph{augmented passing matrix} by $P$.  Here, ``shots" can refer to successful shots, total shots, or a weighted combination of successful shots and missed shots.  Thus, $P$ is a $(M+2) \times (M+2)$ transition probability matrix of a Markov chain with two absorbing states ($S$ and $U$), where, for $1 \leq i, j \leq M$, $p_{i,j}$ is the probability that, if player $i$ has the ball, they pass to player $j$, and $p_{i,S}$ is the probability of transitioning to the $S$ state, and $p_{i,U}$ is the probability of an unsuccessful pass.  Further, $p_{S,S}=p_{U,U}=1$ as $S$ and $U$ are considered terminal states.  Here, each team ball possession is considered a realization of the Markov chain where the state is which player has the ball at that time.  The quantity of interest is
\begin{equation}
    q_j = \P(\text{player } j \text{ is involved in possession } | \text{ possession ends in } S \text{ instead of } U),
\end{equation}
\noindent the \emph{probability that a player was involved in a team possession that ended in a score (or scoring attempt) as opposed to an unsuccessful pass (and/or turnover)}.  The details (including notation) are made rigorous in Appendix \ref{app:p-derivation}.  This probability can be derived in closed form using traditional Markov chain techniques; see Appendix \ref{app:p-derivation}.  Extreme cases provide some quick insight: if $q_j = 1$, it means all shots were filtered through player $j$ in some manner.  Similarly, $q_j=0$ means that player $j$ is not involved in any possession that ended in a shot.  This metric implicitly favors players that contribute by beginning possession but may not be rewarded through in-game stats like goals or assists, since the probability distribution over all team members to begin a possession is utilized in its derivation (see Appendix \ref{app:p-derivation} for details).  Even for strong defensive players, we would still expect $q_j$ to be relatively large, as many shot-based possessions will involve them beginning with the ball.

In order to relate to the $\pi_t$ process, 
\begin{enumerate}
    \item Assume the passing matrix varies over time.  Thus we write $q_{i,t}$ for player $i$'s probability of involvement at time $t$.  Here, $t$ could be per game or season.
    \item Define for each $t$ and player $j$
    \begin{equation}
        \pi_{j,t} = \frac{q_{j,t}}{\sum_{i} q_{i,t}}. \label{eq:pi-defn}
    \end{equation} 
\end{enumerate}
Note here we emphasize the subscript $\pi_{j,t}$ instead of $\pi_t$ as we are comparing across players on the team.  This ensures $0 \leq \pi_{j,t} \leq 1$ and transforms $q_{j,t}$ into a measure of relative player importance. Following Opta, one can also turn this into a running index by using a moving average of the past $K$ games according to $\pi_{j,t_\ell} = \frac{1}{K}\sum_{k=0}^{K-1} \tilde{\pi}_{j,t_{\ell-k}}$ where $\tilde{\pi}_{j,t_{\ell-k}} = \frac{q_{j,t_{\ell-k}}}{\sum_{i} q_{i,t_{\ell-k}}}$ are the game-specific shares according to Equation \eqref{eq:pi-defn} and the games are enumerated at times $t_1, t_2, \cdots$.

\subsection{Illustrative Example for Player Performance Metric}\label{sec:futsal-illustrative-example}

To illustrate this metric, consider a simplified example of Futsal, which is a game of European football played on a smaller court with five player teams.  Denote the players as A, B, C, D, E.  An augmented passing matrix could be as follows

\begin{equation}\label{eq:P-futsal}
P = \left[
\begin{array}{cc}
\left(\begin{array}{ccccc}
0    & 0.40 & 0.25 & 0.20 & 0.10 \\
0.15 & 0    & 0.34 & 0.25 & 0.10 \\
0.05 & 0.15 & 0    & 0.20 & 0.30 \\
0.05 & 0.15 & 0.20 & 0    & 0.20 \\
0    & 0.05 & 0.25 & 0.25 & 0    \\
\end{array}\right) & 
\begin{array}{cc}
0    & 0.05 \\
0.01 & 0.15 \\
0.05 & 0.25 \\
0.10 & 0.20 \\
0.15 & 0.30 \\
\end{array} \\
\begin{array}{cccccccccc}
\hspace{7pt}0\hspace{7pt}&\hspace{7pt} 0 \hspace{7pt}   &\hspace{7pt} 0\hspace{7pt}    &\hspace{7pt} 0\hspace{7pt}    &\hspace{7pt} 0 \hspace{7pt}   \\
\hspace{7pt}0\hspace{7pt}&\hspace{7pt} 0 \hspace{7pt}   &\hspace{7pt} 0\hspace{7pt}    &\hspace{7pt} 0\hspace{7pt}    &\hspace{7pt} 0 \hspace{7pt}   \\
\end{array} & 
\begin{array}{cc}
\hspace{7pt}1\hspace{7pt}    &\hspace{7pt} 0 \hspace{7pt}   \\
\hspace{7pt}0\hspace{7pt}    &\hspace{7pt} 1  \hspace{7pt}  \\
\end{array}
\end{array}
\right]
\end{equation}

\noindent where the upper $5 \times 5$ block is according to the passing matrix.  For example, the first row corresponds to the goalkeeper (A), who has respective probabilities 0.4, 0.25, 0.2, and 0.1 of passing to players B, C, D, E.  The goalkeeper has probability 0 to directly transfer ball possession to the $S$ state (6th column), and has probability 0.05 of making an unsuccessful pass or turnover.  Note that each row sum is 1.  The rows are presented in order of field position, so player B is a defender, players C and D are wing playmakers, and player E is the primary scorer.  Note that the probability of leaving state $S$ is 0, and the same is true for state $U$ (missed passes and/or turnovers), since the team possession ends when the ball reaches that state.  Additionally, let the probability that any given team possession begins with a player be given by initial distribution

\[ \alpha = (0.35, 0.26, 0.17, 0.17, 0.05).\]

\noindent For example, player $B$ begins a team possession with probability 0.26, which could be due to pass interceptions, steals, recoveries, etc.  Denote respectively for players $A, \ldots, E$ as $q_1, \ldots, q_5$, recalling that $q_j$ is the probability that player $j$ is involved in a team possession that ends in state $S$.  

Using the techniques in Appendix \ref{app:p-derivation}, we can determine the vector $(q_1, \ldots, q_5)$ and the standardized version $(\pi_1, \ldots, \pi_5)$.  The results are presented as Scenario (i) in Table \ref{tab:futsal-example}. Player $D$ (right wing) has the largest values, with a probability of $q_4=0.656$ of being involved in a scoring team possession, and a standardized share of $\pi_4=0.232$.  Despite player $E$ having the highest probability of scoring on their own player possession ($0.15$), we have $q_5 = 0.599 < q_4$.  Player $D$'s $q_4$ is larger as more scoring plays were filtered through them.

To illustrate how the metric changes for differing player performance, we consider two additional scenarios, with results presented in Table \ref{tab:futsal-example}:
\begin{itemize}
    \item[(ii)] changing $p_{5,S}=0.25$ from $0.15$ and $p_{5,U}=0.20$ from $0.3$, making player $E$ much more successful at scoring, and
    \item[(iii)] changing $\alpha = (0.30, 0.45, 0.10, 0.10, 0.05)$, making the defender player $B$ begin possessions more often which is true for defensive players (in turn, lowering the initial possession probabilities for $A$, $C, D, E$).
\end{itemize}

\begin{table}[h]
    \centering
    \begin{tabular}{c|ccc|ccc}
        Metric & \multicolumn{3}{|c}{$q_j$} & \multicolumn{3}{|c}{$\pi_j$} \\ \hline
        Scenario & (i) & (ii) & (iii) & (i) & (ii) & (iii)\\ \hline
        Player A & 0.444 & 0.447 & 0.430 & 0.159 & 0.157 & 0.149 \\
        Player B & 0.518 & 0.515 & \textbf{0.644} & 0.185 & 0.181 & \textbf{0.224} \\
        Player C & 0.613 & 0.611 & 0.604 & 0.219 & 0.215 & 0.209 \\
        Player D & \textbf{0.639} & 0.604 & 0.618 & \textbf{0.229} & 0.212 & 0.214 \\
        Player E & 0.579 & \textbf{0.669} & 0.587 & 0.207 & \textbf{0.235} & 0.204 \\
    \end{tabular}
    \caption{Values of $q$ and $\pi$ for Futsal illustrative example.  Scenario (i) corresponds to the default values; (ii) makes player $E$ more successful at scoring compared to missed passes/turnovers, and (iii) significantly increases player $B$'s chances of beginning a possession.  The largest numbers are bolded for readability.}
    \label{tab:futsal-example}
\end{table}

Comparing Scenarios (i) and (ii), player $E$ has $q_5$ increasing from 0.579 to 0.669, an expected increase.  Consequently, $\pi_5=0.235$ compared with $0.207$ as before.  Additionally, player $D$ obtains the largest decrease in their $\pi_4$, with it decreasing by $0.229-0.212 = 0.017$ compared to a decrease of $0.002$ for $A$, and $0.004$ for $B$ and $C$.  This makes sense, as before player $D$ was a source of scoring, but now it is more likely that player $E$ is the one that produces the score.  Looking toward Scenario (iii), player $B$'s $q_2$ skyrockets to 0.644 from 0.518, and $\pi_2$ to 0.224 from 0.185.  Thus, this metric appropriately takes into account defensive opportunities.

We end this example with a few remarks.  First, this section considers only a single $P$ and $\alpha$.  Thus, the values obtained are $q_j$ and $\pi_j$ for a single $t$.  The values will fluctuate as player performance varies from game to game. 

Second, historical data is given in terms of counts, meaning one has the number of passes from each player $i$ to player $j$, their number of scoring shots, unsuccessful shots, and so forth.  A natural way to handle this is to populate a matrix like in \eqref{eq:P-futsal} with the game data, and then standardize each row by its row sum to ensure it is a proper probability transition matrix.  This way, each entry is the empirical probability of that transition for the given game.  Similarly, $\alpha$ can first be populated by the counts of beginning possessions, and then standardized.  Through this procedure, one must formally define what is meant by the $S$ and $U$ states.  In practice, $S$ could be a weighted average of successful shots, shots on-target, and shots off-target.  Similarly, weights for $U$ can be assigned based on the type of turnover.  These weights could be determined through expert opinion and/or existing metrics like Opta \cite{opta}.  Similarly, the values in $\alpha$ could be weighted to assign additional value to certain types of defensive plays like steals.

\section{Workflow for Value Estimation}\label{sec:full-model-usage}

From a practical standpoint, applying the valuation model at contract initialization ($t=0$) to value a player consists of applying the balance Equation \eqref{eq:balance-look-ahead} with $t=0$, summing over the periods of payment.  The framework assumes $t$ is in years, so one should adjust the formula accordingly if one is considering e.g.~weekly paychecks ($h=1/52, 2/52, \cdots)$.  Essentially, this involves a repeated application of Equation \eqref{ctssalary-solved} to obtain $S_{0,1}, S_{0,2}, \cdots, S_{0,T}$.  This is straightforward if one has all of the necessary parameters.  We explain the details, reminding the reader of each parameter's meaning.  This workflow assumes a contract renegotiation for a player already on the team; see Section \ref{sec:transfers} for an adjustment to transfers and free agents.  Thus, full contract valuation is performed according to the following steps:

\begin{enumerate}
    \item Obtain revenue-related parameters $\mu$, $\sigma_R$, and $\rho$, which are respectively the annual revenue growth rate, volatility, and correlation of revenue with the player's performance.  Additionally, one should decide on a risk-free rate $r$ and structure of $(a_1, \ldots, a_T),$ the proportion of revenue paid to players for all times that valuation will occur.
    \item Obtain player-related parameters.  Specific to the $(\pi_t)$ process are their initial performance contribution $\pi_{0}$, long-run performance contribution $\pi^*$, mean reversion parameter $\theta$, and volatility parameter associated with their index $\sigma_{\pi}$.  Additionally one needs their risk premium $\lambda$ (e.g.~due to injuries or violations), which could be estimated with the proportion of games missed during a season due to injuries and red card violations. 
    \item Use Equation \eqref{eq:balance-look-ahead} for all planned payments $(C_1, \ldots, C_T)$, where the $(S_{0,1}, \ldots, S_{0,T})$ are computed using Equation \eqref{ctssalary-solved} and the provided parameters.
\end{enumerate}

\subsection{Illustrative Example of Flexible Contract Valuation}\label{sec:illustrative-example-valuation}

We outline a worked example with assumed parameters for valuation at contract initialization. In this illustrative example, we hope to illuminate flexibility in our method to return a contract term structure satisfactory to both the team and player. Namely, consider a case where a player's agent has asked for a three-year contract where

\begin{itemize}
    \item the salary is front-loaded with year $1$ salary $C_{1}$ triple that of years $2$ and $3$, i.e. $C_{1} = 3C = 3C_{2} = 3C_{3}$, which could act as a signing bonus,
    \item the risk-free rate $r$ is set at the current $3-$yr UK government bond rate of $r=0.0375$,
    \item an agreed-upon risk-premium of $\lambda = 0.01$, and 
    \item a constant fraction of $a_{1} = a_2 = a_3 = 0.5$ of revenue directed towards player salaries over the next three years,
    \item initial revenue $R_0 = 100$ (in \pounds1,000,000) and performance share $\pi_{0} = 0.08$,
    \item long-run performance expectation $\pi^* = 0.10$.
    \item correlation between revenue and performance fraction set at either $(a)$ $\rho = 0$ or $(b)$ $\rho = 0.30$,
    \item corresponding $(R, \pi)$ volatilities fixed at $(\sigma_{R},\sigma_{\pi}) = (0.20,0.25)$, mean reversion rate of $\theta = 4$, and a growth rate $\mu=0.10$ for the team's revenue.
\end{itemize}

Applying Equation \eqref{eq:balance-look-ahead} with $t=0$, the resulting balance equation for the full contract is now

\begin{equation} \label{toyexampleeqn1}
    e^{-r} (3C) + e^{-2r} (C) +  e^{-3r} (C) = e^{-r} S_{0,1} + e^{-2r} S_{0,2} + e^{-3r} S_{0,3}.\\  
\end{equation}

\noindent Plugging in $r = 0.0375$, the year 2 and 3 salaries $C$ can be solved to be

\begin{equation} \label{Ceqn}
    C = \frac{e^{-0.0375} S_{0,1} + e^{-0.0375(2)} S_{0,2} + e^{-0.0375(3)} S_{0,3}}{3 e^{-0.0375} + e^{-0.0375(2)} +  e^{-0.0375(3)}}.
\end{equation}

\noindent Now, Equation \eqref{ctssalary-solved} can be used to find $S_{0,t}$ for $t = 1, 2, 3$ using the parameters described above.


\begin{enumerate}
    \item \textbf{For the case of $\rho = 0$}, one completes the computation in \eqref{Ceqn}, obtaining $C = 3.5349$ (in \pounds1,000,000), resulting in a contract term structure proposal to the player of
        \begin{equation}
          (C_{1},C_{2},C_{3}) = (10.6046, 3.5349, 3.5349).
        \end{equation}
   \item \textbf{For the case of $\rho = 0.30$}, a similar calculation yields $C=3.5745$ so that
\begin{equation}
    (C_{1},C_{2},C_{3}) = (10.7236,3.5745,3.5745).
\end{equation}

\end{enumerate}

We complete our analysis with a of couple observations:

\begin{enumerate}
    \item The salaries $(C_{1},C_{2},C_{3})$ for $\rho = 0.3$ are computed to be higher than if $\rho = 0$. This would coincide with rewarding a player whose increased contributions have a positive effect on team revenue. A general formula for the corresponding ``Greek" $\frac{\partial C}{\partial \rho}$ would be interesting and valuable in player-team negotiations, and is the subject of further research.  A player agent would be expected to argue that their client's performance over the proposed contract has at least some partial (positive) correlation (i.e. $\rho > 0$) with team revenue. 
    \item If $\rho = 0.3$, then the proposed first year salary of \pounds10,723,600 is roughly $9.43\%$ of the total revenue $R_{0}$ at time $0$. The team is committed to offering $50 \%$ of revenue to total team salaries, and so this is a significant share ($\approx \frac{1}{5}$) of total salaries paid for a single player coming off a long-run performance expectation of $10 \%$. However, we also note that team revenue is expected to grow at $10 \%$ with a relatively modest volatility of $20 \%$, and so a team's general manager may in fact be tempted to offer such a deal under these parameter assumptions.
\end{enumerate}

\subsection{Parameter Estimation}

Assume that one has a historic trajectory of the player performance share over a representative period, say $\pi_{t_1}, \ldots, \pi_{t_K}$, which could come from historical game data using the methods described in Section \ref{sec:passing-matrix} or an alternative index like Opta.  One also requires revenue over the same period, $R_{t_1}, \ldots, R_{t_K}$ which should be available to the club in question, and the player should be on the team roster during that period.  The team revenue shares $a_{t_1}, \ldots, a_{t_K}$ should be known and fixed.  In this case, all of the unknown parameters in Equation \eqref{sde1} can be estimated through (approximate) maximum likelihood estimation (MLE) using e.g.~\texttt{R} \cite{R2021programming} with the \texttt{yuima} \cite{iacus2018simulation} package or Python with \texttt{pymle} \cite{kirkby2024pymle}.  One benefit to MLE is that it provides uncertainty estimates of the parameters \cite{casella2024statistical}.  The aforementioned do not straightforwardly handle multi-process dynamics, so we outline an approach using the Euler-Maruyama method \cite{kloeden1992stochastic} in Appendix \ref{app:MLE}. The appendix also outlines the MLE procedure in the single-process case.  We recommend a two-step approach which first estimates the revenue related $\mu$ and $\sigma_R$ and fixes their estimates (which should not vary across players), and then estimates $\theta, \pi^*, \sigma_\pi$, and $\rho$, because $\mu$ and $\sigma_R$ should not vary across team members.

Note that parameters like $\mu$ and $\sigma_R$ are likely known to financial analysts of the club, and may not need to be estimated.  Such expert opinion can generally be used for any parameter, for example if one expects a larger $\pi^*$ than estimated, anticipating player growth (or, varying $\pi^*$ across seasons).



Lastly, the player's risk premium $\lambda$ is independent of the model dynamics. In this case, estimation follows an approach similar to TCV \cite{Tunaru, coluccia2018application}, and is simply the proportion of games a player missed due to injury or violations (e.g.~red card suspensions).


\subsection{Remark on Transfers and Free Agents}\label{sec:transfers}

Valuation of potential incoming players can be performed directly with Equations \eqref{eq:balance-look-ahead} and \eqref{ctssalary-solved} as long as the parameters are known.  Note that $\mu$ and $\sigma_R$ are fixed, so in this case one needs to consider $\pi_0, \pi^*, \theta, \sigma_\pi$, $\rho$, and $\lambda$.  A baseline approach is to provide estimates through expert opinion. Improvements include:
\begin{itemize}
    \item Using the player's historical data to estimate $\lambda$, $\theta$, $\sigma_\pi$, and $\rho$.  This approach makes sense for $\lambda$ if one believes the player's risk premium is the same from their current team.  Similarly, $\sigma_\pi$ is an estimate inherent to player variability, so it makes sense that this parameter should transfer between teams.  On the other hand, $\rho$ is the correlation between the player's performance and revenue for the team performing valuation, so care should be taken as to whether it is reasonable to use the current team's revenue for this estimate.
    \item Similarly, $\pi_0$ and $\pi^*$ could be estimated using their historical data, although in many cases these parameters are likely to change across teams.  One approach is to construct a passing matrix using current team data, and inserting the player into this matrix using analyst estimates for the transition probabilities associated with the player.  This necessitates a critical evaluation of how the player will harmonize with the team, which is expected when a team considers an incoming player.  Along this line of reasoning, one could also consider multiple scenarios for how the player performs to provide a range of estimates.
\end{itemize}

\section{Case Study for European Football}\label{sec:case-study}

This section considers a case study on the EPL (English Premier League). The data covers five seasons spanning 2018--2023.  This period allows analysis of multiple players on the same team with varying contract amounts, which is challenging over a longer time frame due to contract expirations, trades, and retirements.  Each season involves 38 games.  To illustrate the methods presented in this paper, we first examine the dynamics of $\pi_{t}$ across various players and positions using historical game data, exploring how the parameter estimates align with player roles and performance. Subsequently, we demonstrate the player valuation process by comparing historical salaries with the computed $S_{t, T}$. Here, $t$ is measured in years, with $t=0$ corresponding to the beginning of the 2018 season.

The players considered are from Liverpool, Arsenal, and Brighton, representing different positions and roles within their teams.  Specifically,

\begin{itemize}
    \item \textbf{Liverpool}: Mohamed Salah (right winger), the team's primary offensive threat; Trent Alexander-Arnold (right-back), recognized for his creative playmaking; and Virgil Van Dijk (center-back), known for his defensive stability and leadership.  All players were starters for the considered seasons.
    \item \textbf{Arsenal}: Eddie Nketiah (striker), a young goal poacher; Granit Xhaka (central midfielder), a key figure in controlling the game's tempo; and Rob Holding (center-back).  Granit Xhaka was the only consistent starter for all seasons.
    \item \textbf{Brighton}: Pascal Gross (attacking midfielder), the team's creative engine; Solly March (winger/full-back), valued for his versatility and work rate; and Lewis Dunk (center-back), the cornerstone of Brighton's defense.  All teams were starters for the considered seasons.
\end{itemize}

With a variety of teams and consistency of positions, these players provide a practical basis for validating the estimates. The contributions across different positions provide a comprehensive view of $\pi_{t}$ dynamics over time, reflecting both offensive and defensive aspects of the game.  Additionally, it considers players who are not starters as well as a team lacking league superstars (Brighton).

\subsection{Player Performance Share Analysis}

\begin{figure}[!ht]
    \centering
\includegraphics[width=\textwidth, trim={0 1cm 0 0.1cm}, clip]{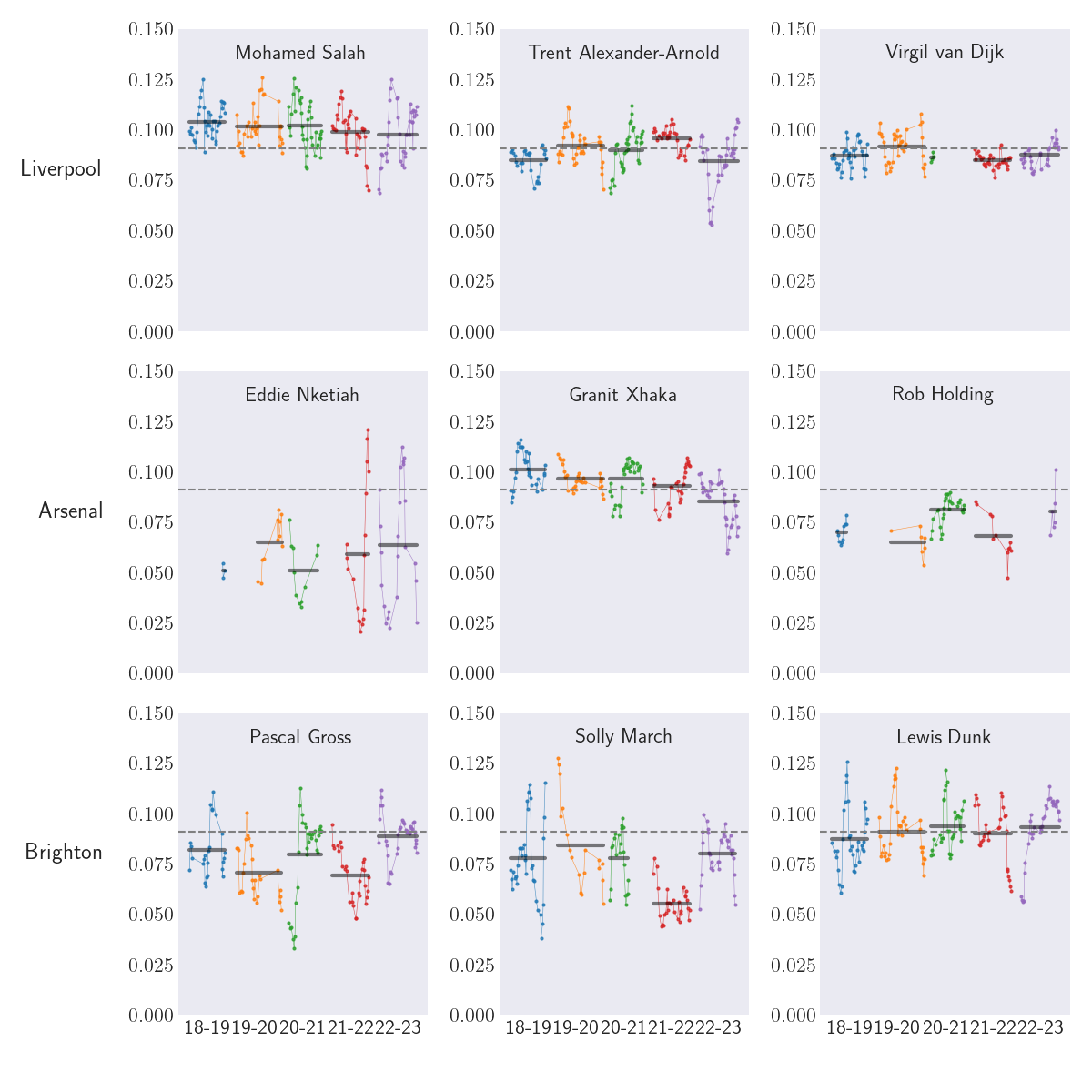}  
    \caption{Calculated $\pi_{t}$ processes using historical game data. The thick dark line is the average over that season, and dashed line is reference of $1/11 \approx 0.0909$, the share if a team consisted of only starters each with equal share.}
    \label{fig:pi-results}
\end{figure}


Data is collected per game\footnote{Game-specific data is from\url{https://www.whoscored.com}} and player shares are calibrated using the estimation methods in Appendix \ref{sec:pi-calibration}.

\begin{table}[!ht]
    \centering
    \begin{tabular}{ll|rrr}
    \toprule
    Team Name & Player Name & \multicolumn{1}{c}{$\hat{\pi}^*$} & \multicolumn{1}{c}{$\hat{\theta}$} & \multicolumn{1}{c}{$\hat{\sigma}_\pi$} \\
    \midrule
    Liverpool & Salah & $0.101$ $\emph{(0.004)}$ & 9.545 $\emph{(0.510)}$ & 0.206 $\emph{(0.011)}$ \\
    Liverpool & Alexander-Arnold & $0.090$ $\emph{(0.006)}$ & $3.595$ $\emph{(0.912)}$ & $0.140$ $\emph{(0.008)}$ \\
    Liverpool & van Dijk & $0.087$ $\emph{(0.002)}$ & $11.31$ $\emph{(0.685)}$ & $0.121$ $\emph{(0.007)}$ \\
    \hline
    Arsenal & Nketiah & $0.046$ $\emph{(0.017)}$ & $3.992$ $\emph{(0.932)}$ & $0.392$ $\emph{(0.034)}$ \\
    Arsenal & Xhaka & $0.093$ $\emph{(0.005)}$ & $3.992$ $\emph{(0.660)}$ & $0.132$ $\emph{(0.008)}$ \\
    Arsenal & Holding & $0.077$ $\emph{(0.006)}$ & $7.271$ $\emph{(2.593)}$ & $0.176$ $\emph{(0.016)}$ \\
    \hline
    Brighton & Gross & $0.077$ $\emph{(0.006)}$ & $4.914$ $\emph{(0.492)}$ & $0.211$ $\emph{(0.012)}$ \\
    Brighton & March & $0.072$ $\emph{(0.008)}$ & $4.704$ $\emph{(0.322)}$ & $0.270$ $\emph{(0.016)}$ \\
    Brighton & Dunk & $0.090$ $\emph{(0.005)}$ & $5.780$ $\emph{(0.663)}$ & $0.208$ $\emph{(0.011)}$ \\
    \bottomrule
    \end{tabular}
    \caption{Estimates related to the $\pi$ process for each player under analysis.  Quantity in parentheses is the standard error of the estimate.  
    }
    \label{tab:player-parameters}
\end{table}

For the players mentioned, we perform calibration steps to obtain their $(\pi_{t})$ trajectory over the five year period. The results are displayed in Figure \ref{fig:pi-results}.  The harvested data is also used to obtain MLE estimates of each player's long-term performance share $\hat{\pi}^*$, rate of reversion $\hat{\theta}$, and volatility parameter $\hat{\sigma}_\pi$.  These are provided in Table \ref{tab:player-parameters}.  Note that one should distinguish between (i) the trajectory of $\pi_{t}$ over time $t$, (ii) the thick dark lines which are the averages over respective seasons, and (iii) $\hat{\pi}^*$, the long-term performance share.

Given that a starting squad comprises 11 players, the $\pi$ values should reasonably align with $0.0909 \approx 1/11$ for starters; this is added to the figure as a dashed line for reference. Note that the $\pi_{t}$ represent proportions over all players involved in the game; thus, $1/11$ is considered a high benchmark. Most starters hover around this benchmark, with Salah consistently outperforming, closer to $0.10$ ($\hat{\pi}^* = 0.101$), Xhaka generally being slightly above ($\hat{\pi}^* = 0.093$), and Gross and March being on the lower end ($\hat{\pi}^* = 0.077$, $0.072$ respectively). Virgil van Dijk is a highly rated center-back, and his season averages hovering near the $1/11$ benchmark is consistent with the discussion in Section~\ref{sec:futsal-illustrative-example}, which illustrated the value in strong defensive play. The only two non-starters (Nketiah, $\hat{\pi}^* = 0.046$, and Holding, $\hat{\pi}^* = 0.077$) tend to be significantly lower than the benchmark as expected, although Nketiah has some periods of stellar performance in the 21--22 and 22--23 seasons. If the contrastingly low values are due to playing time, this suggests Nketiah has value as a starter.

The left column of Figure~\ref{fig:pi-results} displays attackers and offensive threats (Salah, Nketiah, and Gross) who exhibit high variability in their $\pi_{t}$ (respectively, $\hat{\sigma}_{\pi} = 0.206$, $0.392$, and $0.176$). This makes sense as they may be the primary scorer or source of assists in some games but have less involvement in games where shots come from other sources. Brighton's players tend to have higher variability estimates, possibly due to them being a lower-rated team, resulting in more polarized match-ups. On the other hand, Virgil van Dijk shows remarkable consistency both in the figure and his estimated $\hat{\sigma}_{\pi,j} = 0.121$. Xhaka ($\hat{\sigma}_{\pi,j} = 0.132$) and Alexander-Arnold ($\hat{\sigma}_{\pi,j} = 0.140$) are the next most consistent players.

It is important to distinguish between variability that drives the process noise ($\hat{\sigma}_\pi$) and behavior above or below the mean which may be persistent, characterized through $\hat{\theta}$. One way to put $\hat{\theta}$ into perspective is through the estimated autocorrelation (recall equation~\eqref{eq:pi-moments}):
\[
\widehat{\varrho_{\pi}}\left(\frac{h}{52}\right) = \exp\left(-\hat{\theta} \frac{h}{52}\right),
\]
where $h$ is in weeks. Taking the most consistent players mentioned in the last paragraph, with $\hat{\sigma}_\pi$ in the $0.12$--$0.14$ range, one can notice Xhaka and Alexander-Arnold having more ``streaky'' behavior compared to van Dijk, perhaps indicative of hot streaks or slumps. This is reflected in their $\hat{\theta}$ estimates: $3.992$ for Xhaka, $3.595$ for Alexander-Arnold, and $11.31$ for van Dijk. The corresponding correlation between weeks is $\widehat{\varrho_{\pi}}\left(\frac{1}{52}\right)=0.926$ for Xhaka, $0.933$ for Alexander-Arnold, and $0.805$ for van Dijk. At a lag of ten weeks, $\widehat{\varrho_{\pi}}\left(\frac{10}{52}\right)=0.464$, $0.500$, and $0.114$ for the respective players. Essentially, van Dijk more readily returns to his stationary $\hat{\pi}^* = 0.087$, ``forgetting'' highs and lows more quickly compared to Xhaka and Alexander-Arnold, who exhibit more locally persistent trajectories. Nketiah has the same mean reversion $\hat{\theta}$ as Xhaka, and similar local correlation is apparent in his trajectories. Among the other players, the next highest mean reversion strength is observed in Salah with $\hat{\theta} = 9.545$. Note that with Salah's higher $\hat{\sigma}_\pi = 0.206$, the variability appears more ``jumpy,'' where lows are more easily followed by highs and vice versa, both of which have relatively large magnitudes around $\hat{\pi}^* = 0.101$.

The estimated standard errors are mostly consistent across players and show that $\hat{\pi}^*$ is approximately accurate to the third decimal and $\hat{\sigma}_\pi$ to the second. An exception is Nketiah, whose standard errors are generally higher due to sparse data and inconsistent results (see the figure). Holding has a large standard error for $\hat{\theta}$ of $2.593$; otherwise, each player has a standard error of $\hat{\theta}$ between $0.322$ and $0.932$. Accordingly, $\hat{\theta}$ is the hardest parameter to estimate.


\subsection{Player Valuation}\label{section:valuation}

In this section, we illustrate the process of player valuation by comparing players' historic salaries $C_T$, where $T=1, 2, \ldots, 5$, with those computed using equation~\eqref{ctssalary-solved}, denoted as $S_{T-1, T}$, as well as mid-season valuations using $S_{t, T}$ for $T-1 \leq t < T$. Here, $t$ is in years, with $t=0$ representing the beginning of the 2018 season. While financial analysts can, in principle, use granular revenue and salary data if they have access to a team's balance sheet, we can only obtain annual revenue and salary data.\footnote{Salary data is from \url{https://www.spotrac.com}; revenue data is from \cite{Deloitte2019to2023}; game appearance data is from \url{https://www.transfermarkt.com}.} Thus, we have available $R_0, R_1, \ldots, R_4$. Maximum likelihood estimation (MLE) is used to obtain $\hat{\mu}$ and $\hat{\sigma}_R$ using $R_0, \ldots, R_4$. For mid-season valuations, we simulate trajectories of $R_t$ conditional on $R_0, \ldots, R_5$, using the estimated $\hat{\mu}$ and $\hat{\sigma}_R$ for each team, which is feasible since $R_t$ follows a geometric Brownian motion (GBM). All simulations use the same random seed to ensure consistency across players. Note that since $R_0, \ldots, R_4$ are known, the beginning of season valuations $S_{T-1, T}$ for $T = 1, 2, \ldots, 5$ are accurate.  We use the same player estimates for $\hat{\pi}^*$, $\hat{\theta}$, and $\hat{\sigma}_\pi$ as in the previous section.

The parameter $\rho$ represents the correlation between the stochastic fluctuations in the team's revenue $R_t$ and the player's performance index $\pi_t$. Since we only have annual revenue data, we cannot accurately estimate $\rho$ from the data. An attempt was made to replace $R_t$ with its conditional expectation $E[R_t | R_0, \ldots, R_5]$ and perform estimation as if these are realizations from the process. However, this approach is not justified because $\rho$ is specifically meant to capture the correlation between the stochastic components of $R_t$ and $\pi_t$, which are not fully represented in the conditional expectations. Moreover, the resulting standard error estimates were large, indicating unreliable estimates. Therefore, we set $\rho = 0.3$ for all players which is in line with \cite{Tunaru} and \cite{coluccia2018application} and reasonable given the absence of granular revenue data.

We remark that if full revenue data were available, all parameters could be estimated simultaneously using the techniques in Appendix~\ref{app:MLE}. This would allow for a more precise estimation of $\rho$ and a better understanding of the relationship between team revenue and player performance.

Similarly, the player share $a_k$ over season $k$ should practically be known from the club's financial statements. As this data is unavailable, we estimate it as the historic fraction that went out to the club's players over season $k$ for each season.

\begin{table}[ht]
    \centering
    \begin{tabular}{l|rr|rrrrr}
    \toprule
    Team Name & $\hat{\mu}$ & $\hat{\sigma}_R$ & $a_1$ & $a_2$ & $a_3$ & $a_4$ & $a_5$ \\
    \midrule
    Liverpool & 0.073 & 0.116 & 0.219 & 0.233 & 0.255 &  0.208 & 0.239\\
    Arsenal & -0.009 & 0.097 & 0.307 & 0.298 & 0.307 & 0.216 & 0.264\\    
    Brighton & 0.072 & 0.174 & 0.236 & 0.292 & 0.285 & 0.214 & 0.182\\
    \bottomrule
    \end{tabular}
    \caption{Estimates of $\mu$, $\sigma_R$ and calibrated values for each team's player share ($a_k$) process.}
    \label{tab:team-parameters}
\end{table}

\begin{table}[ht]
    \centering
    \begin{tabular}{c|rrr|rrr|rrr}
    \toprule
    Team & \multicolumn{3}{|c}{Liverpool} & \multicolumn{3}{|c}{Arsenal} & \multicolumn{3}{|c}{Brighton}\\
    Player & \multicolumn{1}{c}{Salah} & \multicolumn{1}{c}{Trent} & \multicolumn{1}{c|}{van Dijk} & \multicolumn{1}{c}{Nketiah} & \multicolumn{1}{c}{Xhaka} & \multicolumn{1}{c|}{Holding} & \multicolumn{1}{c}{Gross} & \multicolumn{1}{c}{March} & \multicolumn{1}{c}{Dunk}\\  
    \midrule
    $\hat{\lambda}$ & 0.011 & 0.043 & 0.207 & 0.051 & 0.136 & 0.203 & 0.075 & 0.169 & 0.055\\
    \bottomrule
    \end{tabular}
    \caption{Estimates of risk premium $\hat{\lambda}$ over players.   Trent refers to Trent Alexander-Arnold.}
    \label{tab:risk-premiums}
\end{table}


\subsubsection{Valuation Results}\label{sec:model-application-results}

\begin{figure}[!h]
    \centering
\includegraphics[width=\textwidth, trim={0 1cm 0 0.1cm}, clip]{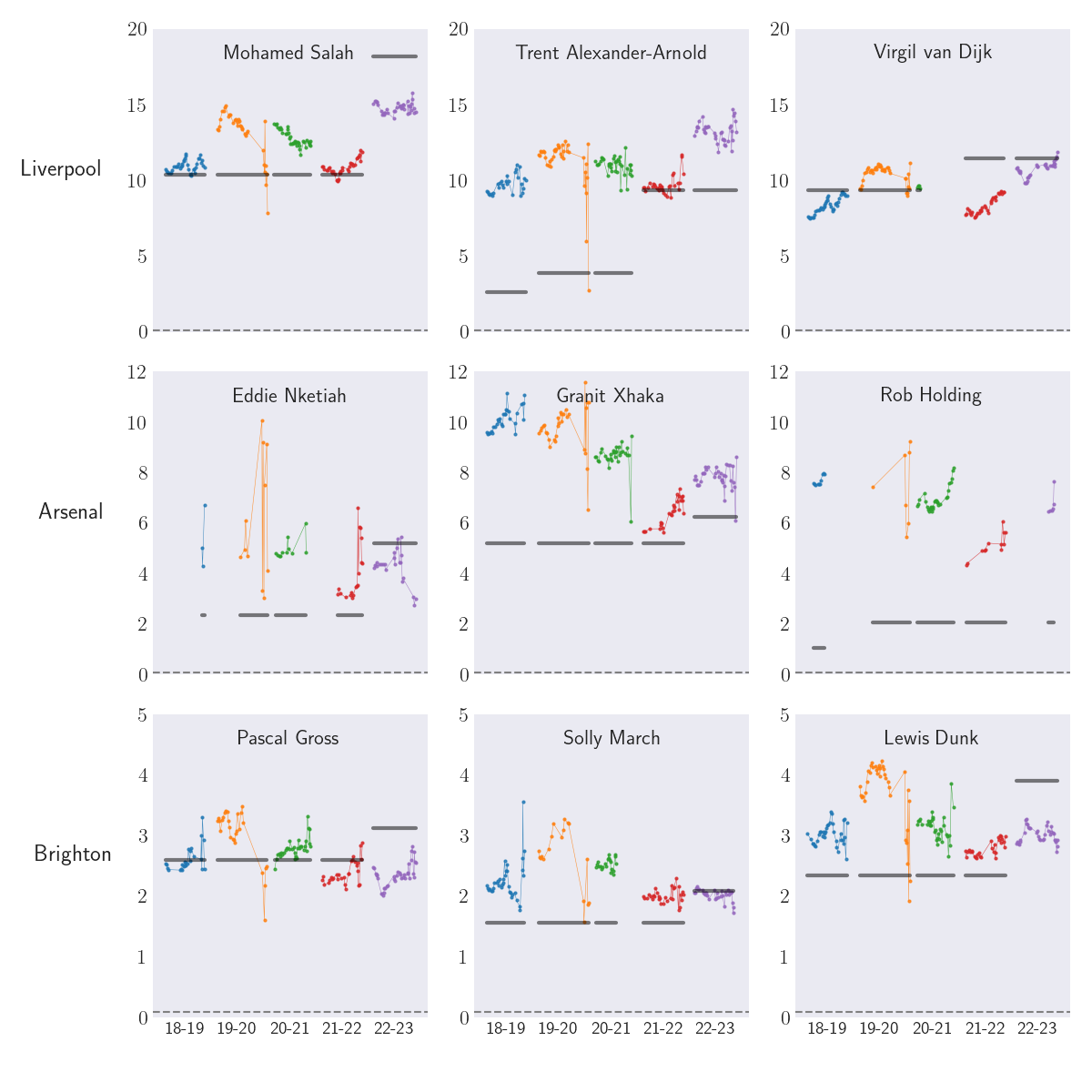}  
    \caption{Time $t$ on $x$-axis, \pounds1,000,000 (annually) on $y$-axis.  Plotted are $C_T$ (thick dark line) and $S_{t,T}$ (colored time series plots), where $T = 1, 2, \ldots, 5$ represents seasons, and $T-1 \leq t < T$ for each season.}
    \label{fig:player-value-plot}
\end{figure}

Our valuation model reveals significant patterns in how player compensation aligns with performance, highlighting both congruencies and discrepancies within current salary structures in European football.

Several players were consistently undervalued relative to their contributions before receiving pay increases. Mohamed Salah's performance data from 2018--2021 suggested a need for a substantial pay increase. He received a significant salary raise in the 2022--2023 season, which addressed his prior undervaluation. However, this increase appears to exceed what his performance metrics justify for that period.  Indeed, his value was higher in 2022--2023 but still below the salary provided—indicating a possible overvaluation. Trent Alexander-Arnold was undercompensated until the 2020--2021 season; his significant pay raise in 2021--2022 aligns with our model's valuation. Yet, he continues to outperform his compensation, suggesting he remains undervalued even after the increase. Granit Xhaka also consistently demonstrated performance exceeding his compensation. The minor salary increase he received in 2022--2023 does not fully reflect his contributions, indicating he is still undervalued according to our model.

Some players received salary increases that our model suggests were unwarranted based on their performance metrics. Lewis Dunk received a substantial pay raise; while his data indicated a need for an increase, the raise may have been larger than justified by his performance. Pascal Gross also received a salary increase for the 2022--2023 season that does not appear warranted, as he underperformed that season according to our model. Conversely, Solly March was slightly underpaid prior to 2022--2023, and his salary increase aligns perfectly with his performance value, suggesting an appropriate adjustment. Eddie Nketiah's situation is nuanced; he received a substantial pay increase in 2022--2023, but its appropriateness is unclear due to his status as a non-starter and limited playing time. Looking at Rob Holding, the model may face limitations when evaluating players with inconsistent playing time, such as Rob Holding and Eddie Nketiah. In particular, the performance index $\pi_t$ ignores games not played due to non-injury, which can distort valuations for non-starters or those frequently absent due to injuries or selection choices. Both players appear to significantly exceed their salary valuations, but this may reflect the model's inability to adjust for limited data. Incorporating additional factors for games not played into the risk premium or modifying the model to account for variability in playing time could improve accuracy for such players.

Players like Virgil van Dijk exhibit a close alignment between compensation and performance. He received a minor salary increase, and his compensation generally matches his performance value throughout the evaluated seasons. His valuation metrics are lower at the beginning of each season and increase toward the end. This trend is due to his higher injury rate and the diminishing impact of the risk premium $\exp(-\lambda(T-t))$ as the season progresses, which reduces the financial effect of potential injuries on his valuation.

These findings underscore the potential of our valuation framework to inform more equitable and performance-based compensation strategies. By identifying discrepancies between player contributions and compensation, clubs can make more informed decisions regarding contract negotiations and salary adjustments. Notably, the first data point in each season corresponds to the valuation $S_{T-1, T}$ which can be directly compared with the actual salary $C_T$.  This comparison provides a clear assessment of how well a player's compensation aligns with their predicted value for the upcoming season, offering valuable insights for aligning salaries with on-field performance.



\section{Discussion and Conclusion}\label{sec:conclusion}

In this paper, we have presented a novel framework for player valuation in European football. Our approach leverages augmented passing matrices modeled as Markov chains to capture the intricate dynamics of player interactions on the field. By utilizing centrality measures derived from these matrices, we quantify individual player influence in a manner that is both dynamic and individualized.

Unlike the findings of Rockerbie and Easton \cite{RockerbieEaston}, who observed in baseball that higher-priced free agents are often overpaid relative to their marginal revenue product (MRP) while lesser-paid free agents offer a bargain, our model appears to work well for both higher- and lower-priced players. Our empirical analysis indicates that the valuation framework accurately captures the contributions of players across different salary levels, providing a balanced assessment that aligns compensation with performance.  While our results sometimes indicate over- or under-compensation, we find many settings where our projected underpayment aligns with a pay raise.

Integrating player performance through the passing matrix offers a unique and robust foundation for capturing on-field performance. By treating the passing matrix as a Markov chain, we can compute closed-form solutions for player involvement probabilities, which serve as the backbone of the financial model's performance index. This approach builds upon the foundational work of Duch et al.\ \cite{duch2010quantifying} and Peña and Touchette \cite{pena2012network}, who demonstrated the effectiveness of network-based analyses in sports performance metrics. By incorporating these methods, we address the limitations of traditional economic models that fail to account for market fluctuations and varied worker productivity \cite{RockerbieEaston}, and improve upon previous stochastic approaches by modeling player performance proportions directly, thus avoiding unrealistic simplifying assumptions.

Empirically validating our model with case studies from Liverpool, Arsenal, and Brighton demonstrates its practical applicability. The calibration of player performance shares using real-world match data confirms that our model effectively captures the nuances of individual contributions across different positions and roles. Our analysis reveals patterns of undervaluation and overvaluation in current compensation structures, highlighting the potential of our framework to inform more equitable and performance-based salary strategies.  This valuation uses the historic estimates for the proportion of revenue allocated to salary ($a_t$), thereby allowing the results to hold regardless of whether the team is win- or profit-maximizing.  We note the absence of a salary cap in European football, which simplifies our modeling approach by removing explicit constraints on team spending. This allows us to model team revenue using geometric Brownian motion without directly incorporating the complexities of constrained optimization \cite{rockerbie2013run}. 

While our discussion focuses on teams in the English Premier League, the methodology can easily extend to other leagues and naturally adapts for use in other sports where player impact can be quantified through a data matrix, such as in hockey. This flexibility underscores the versatility of our approach and its potential for broader application in sports analytics.

Data availability, particularly regarding detailed financial and contractual information, limited our ability to accurately estimate parameters such as $\rho$ and necessitated surrogate choices for mid-season $R_t$. Note, however, that this is not a significant limitation for practitioners within football clubs. Clubs have access to granular revenue data and detailed player statistics, enabling them to apply our methodology with greater precision. The ability to estimate all parameters accurately enhances the practical utility of our model for clubs seeking to optimize their compensation strategies.

In summary, our integration of financial mathematics and network theory provides a significant advancement in the methodology of player valuation in European football. By capturing the stochastic nature of player contributions and their joint dynamics with team revenue, our model offers a robust and flexible framework that bridges the gap between theoretical financial models and practical sports management. By directly modeling player performance proportions and utilizing Markov chains to capture the dynamics of play, we provide a more accurate and realistic representation of player impact. Our findings underscore the potential of our valuation framework to inform more equitable and performance-based compensation strategies. It offers valuable insights for club management in decision-making processes related to contract negotiations, salary adjustments, and performance evaluations.

\subsection{Improvements and Future Directions}

While our work has shown promising results, certain areas warrant further exploration to enhance the model's applicability and accuracy. One aspect is adjusting the model for players with inconsistent playing time. The calibration of the performance index $\pi_t$ currently does not account for games not played. The model requires $\pi_t > 0$, so setting $\pi_t=0$ for those days is infeasible.  A natural remedy is to include an additional premium reflecting the proportion of games not played during a season, similar to the current risk premium estimate that considers injuries and violation-related missed games. Furthermore, it is straightforward to incorporate an annual version of this premium that adjusts for increases in a player's participation level as their career progresses.

Future research could expand the applicability and precision of our model in several key areas. One extension is the incorporation of a salary cap. Integrating a salary cap would necessitate adapting the model to address constrained optimization for optimal salary distribution, providing critical insights for leagues that aim to balance competitive parity with financial sustainability. Additionally, modeling $a_t$ as part of this constrained optimization problem would allow for the simultaneous determination of the total proportion of revenue allocated to salaries and the distribution among individual players. This approach could further elucidate the trade-offs teams face between win-maximizing and profit-maximizing strategies under financial constraints.

Another essential direction involves refining our approach to account for different player roles more comprehensively, starting with goalkeepers. Goalkeepers have a unique role on the field, typically having fewer direct opportunities to influence scoring events, except when initiating possession. This distinction suggests that their contributions might not be fully represented by the current performance metrics. Extending the model to include metrics specifically relevant to goalkeepers, as well as other specialized positions, would ensure more precise player valuations. Moreover, expanding this framework to account for the interconnected dynamics among all team members could further enhance the model’s ability to capture the collective impact on overall team performance. Modeling these interactions through multiple correlated stochastic processes would provide deeper insight into how individual contributions, including specialized roles like goalkeepers, collectively affect team success and revenue.

Investigating parameter sensitivity in contract pricing is also a crucial direction for future study. Analogous to how market volatility impacts financial options, we expect that quantities like $\sigma_\pi$ significantly influence player contract valuations. By analyzing sensitivities such as $\partial / \partial \sigma_\pi$ and other parameters, we can better understand how specific characteristics of player performance impact their financial value, aiding clubs in developing informed contract strategies.

Lastly, applying the model to evaluate the potential contributions of incoming players, such as free agents or transfer targets, is another valuable avenue. Integrating subjective assessments into the passing matrix to predict how new players might fit within the existing team structure would support strategic recruitment. Additionally, analyzing how sensitive the valuation is to these assessments would help ensure that the model remains a robust tool for evaluating new talent effectively.

\bibliography{biblio}

\appendix

\section{Details on Conditional Expectation and Filtrations}\label{app:filtration}

The notion of a filtration is a technical point that allows conditional expectations such as the one in \eqref{eq:conditional-valuation} to be defined properly, and for ``information flow" to be rigorously discussed as what information is available at time $t$. In particular, one can think of $\mathcal{F}_t$ as the information known at time $t$.  This information is not lost, so $\mathcal{F}_s \subseteq \mathcal{F}_t$ for $s < t$, that is, the future holds more information.  This is made rigorous through the concept of a \emph{filtration} $\mathcal{F} = (\mathcal{F}_t)_{t \geq 0}$, the collection of all information sets. The interested reader is referred to \cite{shreve2004stochastic} for more on this technical point. There are two intuitive properties of conditional expectations with respect to filtrations used in this paper:

\begin{itemize}
    \item[(A1)] $\E[\xi_s \eta_s | \mathcal{F}_t] = \xi_s \E[\eta_s | \mathcal{F}_t]$ whenever $\xi_s \in \mathcal{F}_t$.  This means that a stochastic process is treated as deterministic with respect to the $\mathcal{F}_t$--conditional expectation as long as it is known at time $t$.  Intuitively, one can think of $\xi_s$ as being a ``constant'' with respect to $\mathcal{F}_t$, since its value is known.
    \item[(A2)] $\E[\xi_s | \mathcal{F}_0] = \E[\xi_s]$, meaning that the time $t=0$ information is the same as the traditional expectation operator.  Intuitively, this is the situation where there is no additional information known from the stochastic evolution.
\end{itemize}

\section{Calculations for Salary Equation}\label{app:salary-calculation}

Here we outline the derivation of Equation \eqref{ctssalary-solved}.  This involves finding $\E[\pi_{t+h} R_{t+h} | \cF_t]$ and consequently $S_{t,t+h} = e^{-\lambda h}a_{t+h} \E[\pi_{t+h} R_{t+h} | \cF_t]$ follows from Equation \eqref{eq:conditional-valuation}.  Apply It\^o's formula \cite{shreve2004stochastic} to $d(\pi_{t} R_{t})$ using the dynamics in \eqref{sde1}, then apply the conditional expectation operator $\E[\,\cdot\, | \cF_s]$.  Note that $dW$ terms will vanish since $dW_t$ is independent of $\cF_s$. This results in the following equation for $t > s$:

\begin{equation}\label{dt<<1}
\begin{aligned}
 \E[d(\pi_{t} R_{t}) | \mathcal{F}_s] &= (\mu - \theta)\E[\pi_{t} R_{t} | \mathcal{F}_s] dt + \theta\pi^*\E[R_t | \cF_s] dt + \rho\sigma_\pi \sigma_R \E[R_{t} \sqrt{\pi_{t}(1-\pi_{t})} | \cF_s]dt.
\end{aligned}
\end{equation} 

\noindent Now, write $f(t) = \E[\pi_{t} R_{t} | \mathcal{F}_s]$.  Assuming sufficient regularity so that $\E[d\pi_{t} R_{t} | \mathcal{F}_s] = d\E[\pi_{t} R_{t} | \mathcal{F}_s]$ and denoting $f' = \frac{d}{dt} f$, this yields

\begin{equation} \label{ode}
    f'(t) = (\mu - \theta) f(t) + \theta \pi^* R_{s} e^{\mu(t-s)} + c \E[R_t \sqrt{\pi_t (1-\pi_t)} | \cF_s].
\end{equation}




\noindent using the fact that $\E[R_t | \cF_s] = R_{s} e^{\mu(t-s)}$ and denoting $c = \rho \sigma_\pi \sigma_R$.

This is a first-order linear non-homogeneous ordinary differential equation (ODE) with initial condition $f(s) = \pi_s R_s$, which is well studied.  However, the $\E[R_t \sqrt{\pi_t (1-\pi_t)} | \cF_s]$ term makes this difficult to solve.  We use a first order Taylor series approximation of  $v(\pi_t) = \sqrt{\pi_{t} (1-\pi_{t})}$ about its mean reversion parameter $\pi^*:$ 

\begin{equation}
v(\pi_t) = v(\pi^*) + v'(\pi^*)(\pi_t - \pi^*) + \mathcal{O}((\pi_t - \pi^*)^2), \qquad v'(\pi^*) = \frac{1-2\pi^*}{2\sqrt{\pi^*(1-\pi^*)}},
\end{equation}

\noindent which is reasonable since $\pi_t$ is mean-reverting to $\pi^*$.  Truncating the $\mathcal{O}((\pi_t-\pi^*)^2)$ terms and substituting $v(\pi^*) + v'(\pi^*)(\pi_t - \pi^*)$ for $v(\pi_t)$ in $\E[R_t v(\pi_t) | \cF_s]$, the ODE becomes 

\begin{equation} \label{odeappxn}
    f'(t) = \left(\mu - \theta + c v'(\pi^*)\right) f(t) + \left(\theta \pi^* + c v(\pi^*) - c\pi^* v'(\pi^*)\right)R_{s} e^{\mu(t-s)}.
\end{equation}

\noindent This is straightforward to solve and has solution 

\begin{equation}\label{odeappxnsoln}
    f(t) = e^{\mu (t-s)} R_t \left[\pi^* + (\pi_t - \pi^*) e^{(cv'-\theta) (t-s)} + c \sqrt{\pi^*(1-\pi^*)}\frac{1-e^{(cv'-\theta) (t-s)}}{\theta-cv'}\right], \qquad t \geq s.
\end{equation}

Note that this nicely reduces to the initial condition $f(s)=\pi_s R_s$ when plugging in $t=s$.

\subsection{Alternative Methods for Value Calculation}\label{app:salary-alternative}

Denote $f(x, y, t; T) = \E[\pi_T R_T | \pi_t = x, R_t = y] = \E[\pi_T R_T | \cF_t]$.  The Feynman-Kac formula \cite{shreve2004stochastic} gives the partial differential equation
\begin{equation}\label{eq:feynman-pde}
f_t - \theta(x - \pi^*) f_x + \mu y f_y + \frac{1}{2} \sigma_\pi^2 x(1-x)f_{xx} + \frac{1}{2} \sigma_R^2 y^2 f_{yy} + \rho \sigma_\pi \sigma_R y \sqrt{x(1-x)} f_{xy} = 0
\end{equation}
with terminal condition $f(x,y,T) = xy$, using the shorthand $\frac{\partial}{\partial x}f = f_x$, and similar for other partial derivatives.  This can be solved numerically to obtain $\E[\pi_T R_T | \pi_t = x, R_t = y]$ for any $T, t, x$ and $y$ (for example, $f(\pi_0, R_0, 0, T)$).

Alternatively, the dynamics in Equation \eqref{sde1} can be used to obtain a Monte Carlo estimate of $S_{t,t+h}$.  This involves simulating the joint evolution of \( \pi_{t+h} \) and \( R_{t+h} \) given $\cF_t$ a large number of times and computing an average.  The number of simulations $L$ can be increased arbitrarily to decrease the standard error of the Monte Carlo estimate (which scales in $1/\sqrt{L}$ \cite{glasserman2004monte}).  However, the joint pdf is unknown so there is an additional time-step discretization error.

To illustrate the accuracy of our method, we use the setup in Example \ref{sec:illustrative-example-valuation} and compare our first-order approximation \( S_{0,T} \) as in Equation \eqref{ctssalary-solved} with a zeroth-order approximation \( S^{0}_{0,T} \) (obtained by replacing \( \pi_t \) with \( \pi^* \) in the approximation). We also compare these with a numerical PDE estimate \( S^{\text{PDE}}_{0,T} \) based on Equation \eqref{eq:feynman-pde} and a Monte Carlo estimate \( S^{\text{MC}}_{0,T} \) for \( T = \frac{1}{50}, 1, 2, 3 \). The short time horizon \( T = \frac{1}{50} \) (approximately one week) is included to assess approximation quality over shorter durations.

The PDE estimate is obtained using Wolfram Mathematica 13.3 with default settings, a minimum grid size of 100, and accuracy and precision goals set to 6. The Monte Carlo estimate employs the \texttt{yuima} package in \texttt{R} \cite{iacus2018simulation} with \( M = 10^6 \) independent sample paths of \( \pi_{t} \) and \( R_{t} \), using a timestep of \( \Delta t = 0.001 \). Table \ref{tab:MC-results} presents the results.

\begin{table}[h]
    \centering
    \begin{tabular}{c|rrrr|r}
        \hline
        \( T \) & \( S_{0,T} \) & \( S^{0}_{0,T} \) & \( S^{\text{PDE}}_{0,T} \) & \( S^{\text{MC}}_{0,T} \) & \( \text{SE}(S^{\text{MC}}_{0,T}) \) \\ \hline
        \( \dfrac{1}{50} \) & 4088.192 & 4088.561 & 4088.162 & 4083.497 & 0.4994 \\
        1                   & 5511.127 & 5511.251 & 5510.577 & 5507.710 & 2.0395 \\
        2                   & 6053.327 & 6053.006 & 6052.819 & 6051.692 & 2.5964 \\
        3                   & 6623.869 & 6623.499 & 6623.427 & 6621.244 & 3.1936 \\
        \hline
    \end{tabular}
    \caption{Comparing valuation approximations for \( T = \dfrac{1}{50}, 1, 2, 3 \).}
    \label{tab:MC-results}
\end{table}

First, observe that the Monte Carlo standard errors increase with \( T \), reflecting greater variability over longer time horizons. This increase underscores the inclusion of the short-term horizon \( T = 1/50 \), where the standard error is relatively low (0.4994), allowing for a more precise comparison. The first-order \( S_{0,T} \) is in close agreement with \( S^{\text{PDE}}_{0,T} \), differing by only 0.03. In contrast, the zeroth-order \( S^{0}_{0,T} \) differs from \( S_{0,T} \) by 0.369 and from \( S^{\text{PDE}}_{0,T} \) by 0.399, indicating that including the first-order term significantly enhances the approximation's accuracy.


At longer time horizons (\( T = 2 \) and \( T = 3 \)), an interesting pattern emerges: the PDE estimates \( S^{\text{PDE}}_{0,T} \) are closer to the zeroth-order \( S^{0}_{0,T} \) than to the first-order \( S_{0,T} \).   This reversal is unexpected, as the first-order approximation is an objective improvement over the zeroth-order approximation. A likely explanation is that the PDE solutions at longer time horizons suffer from discretization inaccuracies. Despite using a finer grid and higher accuracy settings than the default, these inaccuracies gave estimates closer to the zeroth-order approximation.  Hence, the first-order $S_{0,T}$ is doing quite well.

The Monte Carlo estimates have been mostly excluded from this analysis as the $T=1/50$ case reveals a clear issue with discretization error, and longer periods will lead to more significant biases.  Additionally, the standard errors increase, effectively reducing the accuracy of the estimates.



\section{On Passing Matrix Markov Chains}

\subsection{Derivation of Markov Chain Probability}\label{app:p-derivation}

Let $(X_n)_{n=0}^\infty$ represent the \emph{team possession process}, indicating which player has the ball at step $n$ for a given team possession, governed by the Markov chain transition probabilities $P$ \cite{grimmett2020probability}, where a ``step'' refers to a transition in ball possession (pass, shot, etc.). In addition, we treat the initial distribution $\alpha_i = \P(X_0 = i)$ as the probability of player $i$ beginning a team possession (through start of half, steal, penalty, etc.). From this Markov chain, the metric of interest is

\begin{equation}
    q_i = \P(A_i | X_{\infty} = S), \qquad A_i = \{X_\ell = i \text{ for some } \ell = 0, 1, \ldots\}
\end{equation} 

Here, $X_{\infty} = S$ means the team possession ended in $S$ instead of $U$, and $A_i$ is the event that player $i$ had the ball at some point during that team possession.  Note that $P$ is a finite Markov chain.  For simplicity, assume that all states communicate with one another except for the two absorbing states $S$ and $U$.  It is well known that one can find the probability of absorption into one of these states starting in state $j$. Now,

\begin{align*}
    q_i = \P(A_i | X_{\infty} = S) &= \frac{\P(A_i \cap \{X_{\infty} = S\})}{\P(X_{\infty} = S)}.
\end{align*}

\noindent For the numerator

\begin{align*}
    \P(A_i \cap \{X_{\infty} = S\}) &= \sum_j \P(A_i \cap \{X_{\infty} = S\} | X_0 = j)\P(X_0=j)\\
        &= \sum_j \left[ \P(X_{\infty} = S | X_0 = j) - \P(\{X_{\infty} = S\} \backslash A_i | X_0 = j)\right] \P(X_0=j)
\end{align*}

\noindent where $\{X_{\infty} = S\} \backslash A_i$ is the set difference, i.e.~the case of absorption into $S$ but player $i$ never having possession.  This probability can be computed by considering a new Markov chain in which state $i$ is treated as absorbing, and by determining the probability of reaching state $S$ before reaching state $i$ (refer to \cite{grimmett2020probability} for more details).  For the denominator,

\begin{align*}
    \P(X_{\infty} = S) = \sum_j \P(X_{\infty} = S | X_0 = j)\P(X_0 = j)
\end{align*}

Note that the initial distribution $\alpha_j = \P(X_0=j)$ appears in both the numerator and denominator, weighing accordingly to the probability that player $j$ begins a team possession.

\section{Likelihoods and Estimation}\label{app:MLE}

Considering observed trajectories of the univariate processes $(\pi_{t_1}, \ldots, \pi_{t_N})$ and $(R_{t_1}, \ldots, R_{t_N})$, maximum likelihood estimation (MLE) \cite{casella2024statistical} estimates their unknown parameters with the goal to maximize the probability density functions of the observations:
\begin{align}\label{eq:MLE-pi}
    \hat{\pi}^*, \hat{\theta}, \hat{\sigma}_\pi &= \text{argmax}_{\pi^*, \theta, \sigma_\pi} f(\pi_{t_1}, \ldots, \pi_{t_N} ; \pi^*, \theta, \sigma_\pi),
\end{align}
and similarly for $\mu, \sigma_R$.  Since the processes in question are Markov, one can instead write (again, in the $\pi$ case)
\[f(\pi_{t_1}, \ldots, \pi_{t_N} ; \pi^*, \theta, \sigma_\pi) = f(\pi_1; \pi^*, \theta, \sigma_\pi) \prod_{i=2}^N f(\pi_{t_i} | \pi_{t_{i-1}} ; \pi^*, \theta, \sigma_\pi)\]
where the $f(\pi_{t_i} | \pi_{t_{i-1}} ; \pi^*, \theta, \sigma_\pi)$ are the transition densities and $f(\pi_1; \pi^*, \theta, \sigma_\pi)$ is the stationary pdf.  Since $R$ is a GBM, the MLE's for $\mu$ and $\sigma_R$ are available in closed form
\begin{equation}\label{eq:R-MLE}
    \hat{\sigma}^2_R = \frac{1}{N} \sum_{i=1}^N\left(\log\left(R_{t_i} / R_{t_{i-1}}\right) - \hat{y}\right)^2, \qquad \hat{\mu} = \frac{1}{2} \hat{\sigma}_R^2 + \hat{y}
\end{equation}
where $\hat{y} = \frac{1}{N}\sum_{i=1}^N \log\left(R_{t_i} / R_{t_{i-1}}\right)$.  For $\hat{\pi}^*, \hat{\theta}$ and $\hat{\sigma}_\pi$, the transition densities $f(\pi_{t_i} | \pi_{t_{i-1}} ; \pi^*, \theta, \sigma_\pi)$ have no closed form.  The Python package \texttt{pymle} \cite{kirkby2024pymle} supports several approximations and obtains the MLEs.  Our analysis used the Kessler \cite{kessler1997estimation} density estimate which assumes a Gaussian transition density using a second order mean and variance approximation according to dynamics in \eqref{sde1}.

\subsubsection*{Bivariate Dynamics} Estimating $\rho$ in addition to other parameters involves the maximization
\begin{align}\label{eq:MLE}
    \hat{\mu}, \hat{\sigma}_R, \hat{\pi}^*, \hat{\theta}, \hat{\sigma}_\pi, \hat{\rho} &= \text{argmax}_{\mu, \sigma_R, \pi^*, \theta, \sigma_\pi, \rho} \prod_{i=1}^N f(\pi_{t_i}, R_{t_i} | \pi_{t_{i-1}}, R_{t_{i-1}} ; \mu, \sigma_R, \pi^*, \theta, \sigma_\pi, \rho).
\end{align}
Note that some of the estimates may be first fixed, for example according to a univariate process estimate as was done in the paper.  In this case one simply plugs in that estimate prior to maximizing (e.g.~$\mu = \hat{\mu}$ with \eqref{eq:R-MLE}).  The aforementioned package does not handle multi-process dynamics, so we outline an approach using the Euler-Maruyama method \cite{kloeden1992stochastic}, similar to the Kessler method.  Working with $L_t = \log(R_t)$ is easier, in which case Ito's lemma shows $dL_t = (\mu - \sigma_R^2/2)dt + \sigma_R dW_t^R$.  Euler-Maruyama discretizes the dynamics in \eqref{sde1} so that
\begin{align}
    \pi_{t_i} &= \pi_{t_{i-1}} - \theta (\pi_{t_{i-1}}-\pi^*)h_i + \sigma_\pi \sqrt{\pi_{t_{i-1}}(1-\pi_{t_{i-1}})}Z_{i}^\pi,\\ 
    L_{t_i} &= L_{t_{i-1}} + (\mu - \sigma_R^2/2)h_i + \sigma_R Z_{i}^R,\\
\end{align}
where $h_i = t_i - t_{i-1}$, and the $Z_i^\pi, Z_i^R$ come from the Brownian motion innovations.  These are normally distributed with mean zero and variance $h_i$, are independent of $Z_j^\pi, Z_j^R$ for $i \neq j$, and have $\text{corr}(Z_i^\pi, Z_i^R)=\rho$.  Consequently, the bivariate transition densities $f(\pi_{t_i}, L_{t_i} | \pi_{t_{i-1}}, L_{t_{i-1}})$ are bivariate normal with mean vector and covariance matrix
\begin{equation}
    \begin{bmatrix} \pi_{t_{i-1}} - \theta (\pi_{t_{i-1}}-\pi^*)h_i \\
    L_{t_{i-1}} + (\mu - \sigma_R^2/2)h_i \end{bmatrix}, \qquad
    \begin{bmatrix}
    \sigma^2_\pi\pi_{t_{i-1}}(1-\pi_{t_{i-1}})h_i & \rho \sigma_R \sigma_\pi\sqrt{\pi_{t_{i-1}}(1-\pi_{t_{i-1}})} h_i \\
    \rho \sigma_R \sigma_\pi\sqrt{\pi_{t_{i-1}}(1-\pi_{t_{i-1}})} h_i & \sigma_R^2 h_i\end{bmatrix}.
\end{equation}
Here, $\sigma_R, \sigma_\pi, \theta > 0$, $-1 \leq \rho \leq 1$, $0 \leq \pi^* \leq 1$, and $\min(\pi^*, 1-\pi^*) \geq \sigma^2_\pi / (2\theta)$.  These are used in tandem with Equation \eqref{eq:MLE} to estimate any unknown parameters.

\subsection{Estimation of Player Performance Shares}\label{sec:pi-calibration}

For a game that occurred at time $t$, the procedure to obtain $\pi_{j,t}$ for all players $j = 1, \ldots, M$ on the team is as follows:

\begin{enumerate}
    \item Construct the passing frequency matrix, i.e.~the $M \times M$ matrix where the $i,j$ entry is the number of passes from player $i$ to player $j$.  Denote the $i,j$ entry of this matrix as $n_{i,j}$.
    \item Add a row and column associated with the $S$ and $U$ state.  The added rows should be zeroes except for the diagonal entries which equal 1.
    \item The $i$th entry of the $S$ column is a convex combination of the number of shots made and shots missed for player $i$:
    \begin{equation}\label{eq:n-i-s}
        n_{i,S} = w_S \cdot n_{i,\text{score}} + (1-w_S) \cdot n_{i,\text{miss}},
    \end{equation}
    where $n_{i,\text{score}}$ is the number of times player $i$ scored (implicitly for the game at time $t$), and $n_{i,\text{miss}}$ is the number of missed shots.  We follow the weighting scheme of Opta \cite{opta} with $w_S = 5/6$, which counts a missed shot as 20\% of a score.
    \item The $i$th entry of the $U$ column is the number of missed passes by that player.
    \item Divide each row of the resulting $(M+2) \times (M+2)$ matrix by its row sum to obtain $P_t$, the empirical augmented passing matrix for the game at time $t$.
    \item Let $\alpha_t \in \mathbb{R}^M$ be the vector whose $i$th entry, $i = 1, \ldots, M$, is the proportion of times player $i$ began possession for the game at time $t$.
    \item Use the methods described in Appendix \ref{app:p-derivation} to obtain $q_{j,t}, j = 1, \ldots, M$, and consequently use Equation \eqref{eq:pi-defn} to obtain $\pi_{j,t}, j = 1, \ldots, M$, using a six-game moving average.
\end{enumerate}

We remark that step 4 could be adjusted to include turnovers, or even a weighted approach like in \eqref{eq:n-i-s} to more heavily penalize egregious mistakes.  Step 6 could also use a weighting scheme, e.g.~to add weight to steals.

\end{document}